%% file: nat.tex
\documentclass[10pt,twocolumn]{article}
\usepackage{fullpage}
\usepackage{epsfig}
\usepackage{times}
\usepackage[compact]{titlesec}

\long\def\com#1{}

\title{Peer-to-Peer Communication Across Network Address Translators}
\author{
	Bryan Ford \\
	Massachusetts Institute of Technology \\
	{\tt baford@mit.edu}
\and
	Pyda Srisuresh \\
	Caymas Systems, Inc. \\
	{\tt srisuresh@yahoo.com} \\
\and
	Dan Kegel \\
	{\tt dank@kegel.com}
}
\date{}

\begin{document}
\maketitle

\begin{quote}
{\em 
J'fais des trous, des petits trous$\dots$ \\
toujours des petits trous} \\
\centerline{- S. Gainsbourg}
\end{quote}

\input{abs}
\input{intro}

\input{trav}

\input{udp}

\input{tcp}

\input{good}

\input{eval}

\input{related}

\input{conc}

\begin{footnotesize}
\bibliography{nat}
\bibliographystyle{plain}
\end{footnotesize}

\end{document}

%% file: abs.tex
\begin{abstract}
Network Address Translation (NAT)
causes well-known difficulties
for peer-to-peer (P2P) communication,
since the peers involved may not be reachable
at any globally valid IP address.
Several NAT traversal techniques are known,
but their documentation is slim,
and data about their robustness or relative merits is slimmer.
This paper documents and analyzes
one of the simplest but most robust and practical
NAT traversal techniques,
commonly known as ``hole punching.''
Hole punching is moderately well-understood for UDP communication,
but we show how it can be reliably used 
to set up peer-to-peer TCP streams as well.
After gathering data on the reliability of this technique
on a wide variety of deployed NATs,
we find that about 82\% of the NATs tested support hole punching for UDP,
and about 64\% support hole punching for TCP streams.
As NAT vendors become increasingly conscious of
the needs of important P2P applications
such as Voice over IP and online gaming protocols,
support for hole punching is likely to increase in the future.
\end{abstract}

%% file: intro.tex
\section{Introduction}
\label{sec-intro}

\begin{figure}[t]
\centerline{\epsfig{file=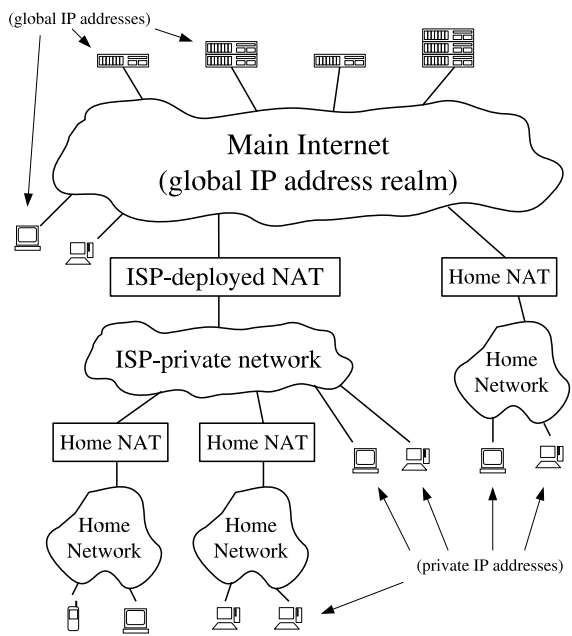, scale=0.40}}
\caption{Public and private IP address domains}
\label{fig-twicenat}
\end{figure}

The combined pressures of tremendous growth
and massive security challenges
have forced the Internet to evolve
in ways that make life difficult for many applications.
The Internet's original uniform address architecture,
in which every node has a globally unique IP address
and can communicate directly with every other node,
has been replaced with a new {\em de facto} Internet address architecture,
consisting of a global address realm
and many private address realms
interconnected by Network Address Translators (NAT).
In this new address architecture,
illustrated in Figure~\ref{fig-twicenat},
only nodes in the ``main,'' global address realm
can be easily contacted from anywhere in the network,
because only they have unique, globally routable IP addresses.
Nodes on private networks
can connect to other nodes on the same private network,
and they can usually open TCP or UDP connections
to ``well-known'' nodes in the global address realm.
NATs on the path
allocate temporary public endpoints for outgoing connections,
and translate the addresses and port numbers
in packets comprising those sessions,
while generally blocking all incoming traffic
unless otherwise specifically configured.

The Internet's new {\em de facto} address architecture
is suitable for client/server communication
in the typical case when the client is on a private network
and the server is in the global address realm.
The architecture makes it difficult for two nodes
on {\em different} private networks to contact each other directly,
however,
which is often important
to the ``peer-to-peer'' communication protocols
used in applications such as teleconferencing and online gaming.
We clearly need a way to make such protocols
function smoothly in the presence of NAT.

One of the most effective methods
of establishing peer-to-peer communication
between hosts on different private networks
is known as ``hole punching.''
This technique
is widely used already in UDP-based applications,
but essentially the same technique also works for TCP.
Contrary to what its name may suggest,
hole punching does not compromise the security of a private network.
Instead,
hole punching enables applications to function
{\em within} the the default security policy of most NATs,
effectively signaling to NATs on the path
that peer-to-peer communication sessions
are ``solicited''
and thus should be accepted.
This paper documents hole punching for both UDP and TCP,
and details the crucial aspects of both application and NAT behavior
that make hole punching work.

Unfortunately,
no traversal technique works with all existing NATs,
because NAT behavior is not standardized.
This paper presents some experimental results
evaluating hole punching support in current NATs.
Our data is derived from results submitted by users throughout the Internet
by running our ``NAT Check'' tool
over a wide variety of NATs by different vendors.
While the data points were gathered from a ``self-selecting'' user community
and may not be representative
of the true distribution of NAT implementations deployed on the Internet,
the results are nevertheless generally encouraging.

While evaluating basic hole punching,
we also point out variations that can make hole punching work
on a wider variety of existing NATs
at the cost of greater complexity.
Our primary focus, however,
is on developing the {\em simplest} hole punching technique
that works cleanly and robustly
in the presence of ``well-behaved'' NATs
in any reasonable network topology.
We deliberately avoid excessively clever tricks
that may increase compatibility
with some existing ``broken'' NATs in the short term,
but which only work some of the time
and may cause additional unpredictability and network brittleness
in the long term.

Although the larger address space of IPv6~\cite{rfc2460}
may eventually reduce the need for NAT,
in the short term IPv6
is {\em increasing} the demand for NAT,
because NAT itself provides the easiest way to achieve interoperability
between IPv4 and IPv6 address domains~\cite{rfc2766}.
Further,
the anonymity and inaccessibility of hosts on private networks
has widely perceived security and privacy benefits.
Firewalls are unlikely to go away even when there are enough IP addresses:
IPv6 firewalls
will still commonly block unsolicited incoming traffic by default,
making hole punching useful even to IPv6 applications.

The rest of this paper is organized as follows.
Section~\ref{sec-trav}
introduces basic terminology and NAT traversal concepts.
Section~\ref{sec-udp}
details hole punching for UDP,
and Section~\ref{sec-tcp}
introduces hole punching for TCP.
Section~\ref{sec-good}
summarizes important properties a NAT must have
in order to enable hole punching.
Section~\ref{sec-eval}
presents our experimental results
on hole punching support in popular NATs,
Section~\ref{sec-related} discusses related work,
and Section~\ref{sec-conc} concludes.

%% file: trav.tex
\section{General Concepts}
\label{sec-trav}

This section introduces basic NAT terminology
used throughout the paper,
and then outlines general NAT traversal techniques
that apply equally to TCP and UDP.

\subsection{NAT Terminology}

This paper adopts the NAT terminology and taxonomy
defined in RFC 2663~\cite{rfc2663},
as well as additional terms defined more recently in RFC 3489~\cite{rfc3489}.

Of particular importance is the notion of session.
A {\em session endpoint} for TCP or UDP
is an (IP address, port number) pair,
and a particular {\em session} is uniquely identified by
its two session endpoints.
From the perspective of one of the hosts involved,
a session is effectively identified by the 4-tuple
(local IP, local port, remote IP, remote port).
The {\em direction} of a session
is normally the flow direction of the packet
that initiates the session:
the initial SYN packet for TCP,
or the first user datagram for UDP.

Of the various flavors of NAT,
the most common type is {\em traditional} or {\em outbound} NAT,
which provides an asymmetric bridge
between a private network and a public network.
Outbound NAT by default allows only outbound sessions to traverse the NAT:
incoming packets are dropped unless the NAT identifies them
as being part of an existing session
initiated from within the private network.
Outbound NAT conflicts with peer-to-peer protocols
because when both peers desiring to communicate are ``behind''
(on the private network side of) two different NATs,
whichever peer tries to initiate a session,
the other peer's NAT rejects it.
NAT traversal entails
making P2P sessions look like ``outbound'' sessions
to {\em both} NATs.

Outbound NAT has two sub-varieties:
{\em Basic NAT}, which only translates IP addresses,
and {\em Network Address/Port Translation} (NAPT),
which translates entire session endpoints.
NAPT, the more general variety,
has also become the most common
because it enables the hosts on a private network
to share the use of a {\em single} public IP address.
Throughout this paper we assume NAPT,
though the principles and techniques we discuss
apply equally well (if sometimes trivially) to Basic NAT.

\subsection{Relaying}

\begin{figure}[t]
\centerline{\epsfig{file=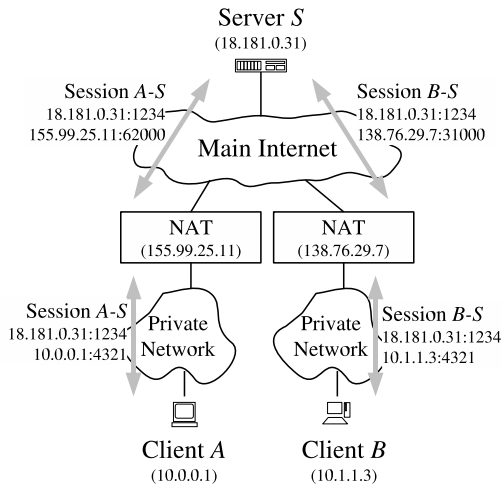, scale=0.40}}
\caption{NAT Traversal by Relaying}
\label{fig-relay}
\end{figure}

The most reliable---but least efficient---method
of P2P communication across NAT
is simply to make the communication
look to the network like standard client/server communication,
through relaying.
Suppose two client
hosts $A$ and $B$
have each initiated TCP or UDP 
connections to a well-known server $S$,
at $S$'s global IP address 18.181.0.31 and port number 1234.
As shown in Figure~\ref{fig-relay},
the clients reside on separate private networks, and
their respective NATs prevent either client from directly
initiating a connection to the other.
Instead of attempting a direct connection, the two clients can simply
use the server $S$ to relay messages between them.  For example, to
send a message to client $B$, client $A$ simply sends the message to
server $S$ along its already-established client/server connection, and
server $S$ forwards the message on to client $B$ using its existing
client/server connection with $B$.

Relaying always works
as long as both clients can connect to the server.
Its disadvantages are that
it consumes the server's processing power and network bandwidth,
and communication latency between the peering clients
is likely increased even if the server is well-connected.
Nevertheless,
since there is no more efficient technique
that works reliably on all existing NATs,
relaying is a useful fall-back strategy
if maximum robustness is desired.
The TURN protocol~\cite{rosenberg03traversal}
defines a method of implementing relaying
in a relatively secure fashion.

\subsection{Connection Reversal}

\begin{figure}[t]
\centerline{\epsfig{file=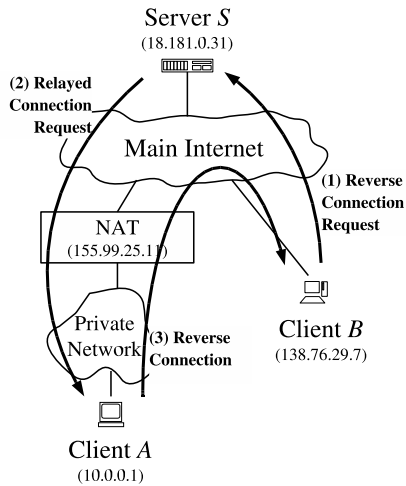, scale=0.40}}
\caption{NAT Traversal by Connection Reversal}
\label{fig-reversal}
\end{figure}

Some P2P applications use
a straightforward but limited technique,
known as {\em connection reversal},
to enable communication
when both hosts have connections to a well-known rendezvous server $S$
and only one of the peers is behind a NAT,
as shown in Figure~\ref{fig-reversal}.
If $A$ wants to initiate a connection to $B$,
then a direct connection attempt works automatically,
because $B$ is not behind a NAT
and $A$'s NAT interprets the connection as an outgoing session.
If $B$ wants to initiate a connection to $A$, however,
any direct connection attempt to $A$ is blocked by $A$'s NAT.
$B$ can instead relay a connection request to $A$
through a well-known server $S$,
asking $A$ to attempt a ``reverse'' connection back to $B$.
Despite the obvious limitations of this technique,
the central idea of using a well-known rendezvous server
as an intermediary to help set up direct peer-to-peer connections
is fundamental to the more general hole punching techniques
described next.


%% file: udp.tex
\section{UDP Hole Punching}
\label{sec-udp}

UDP hole punching enables two clients to set up
a direct peer-to-peer UDP session
with the help of a well-known rendezvous server,
even if the clients are both behind NATs.
This technique was mentioned
in section 5.1 of RFC 3027~\cite{rfc3027},
documented more thoroughly elsewhere on the Web~\cite{kegel99nat},
and used in recent
experimental Internet protocols~\cite{rosenberg03ice, huitema04teredo}.
Various proprietary protocols,
such as those for on-line gaming,
also use UDP hole punching.

\begin{figure*}[t]
\centerline{\epsfig{file=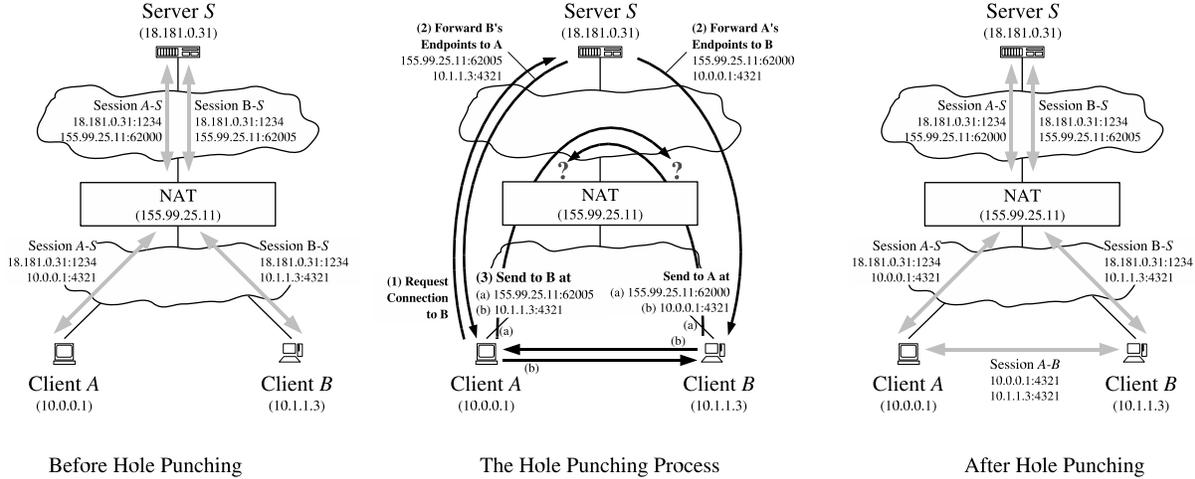, scale=0.34}}
\caption{UDP Hole Punching, Peers Behind a Common NAT}
\label{fig-samenat}
\end{figure*}

\begin{figure*}[t]
\centerline{\epsfig{file=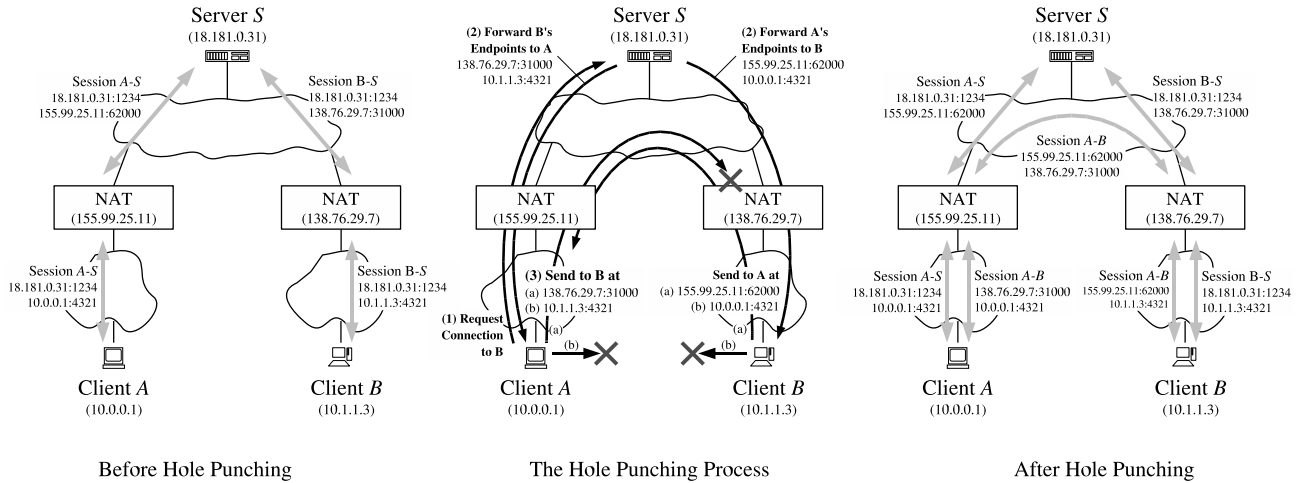, scale=0.34}}
\caption{UDP Hole Punching, Peers Behind Different NATs}
\label{fig-diffnat}
\end{figure*}

\begin{figure*}[t]
\centerline{\epsfig{file=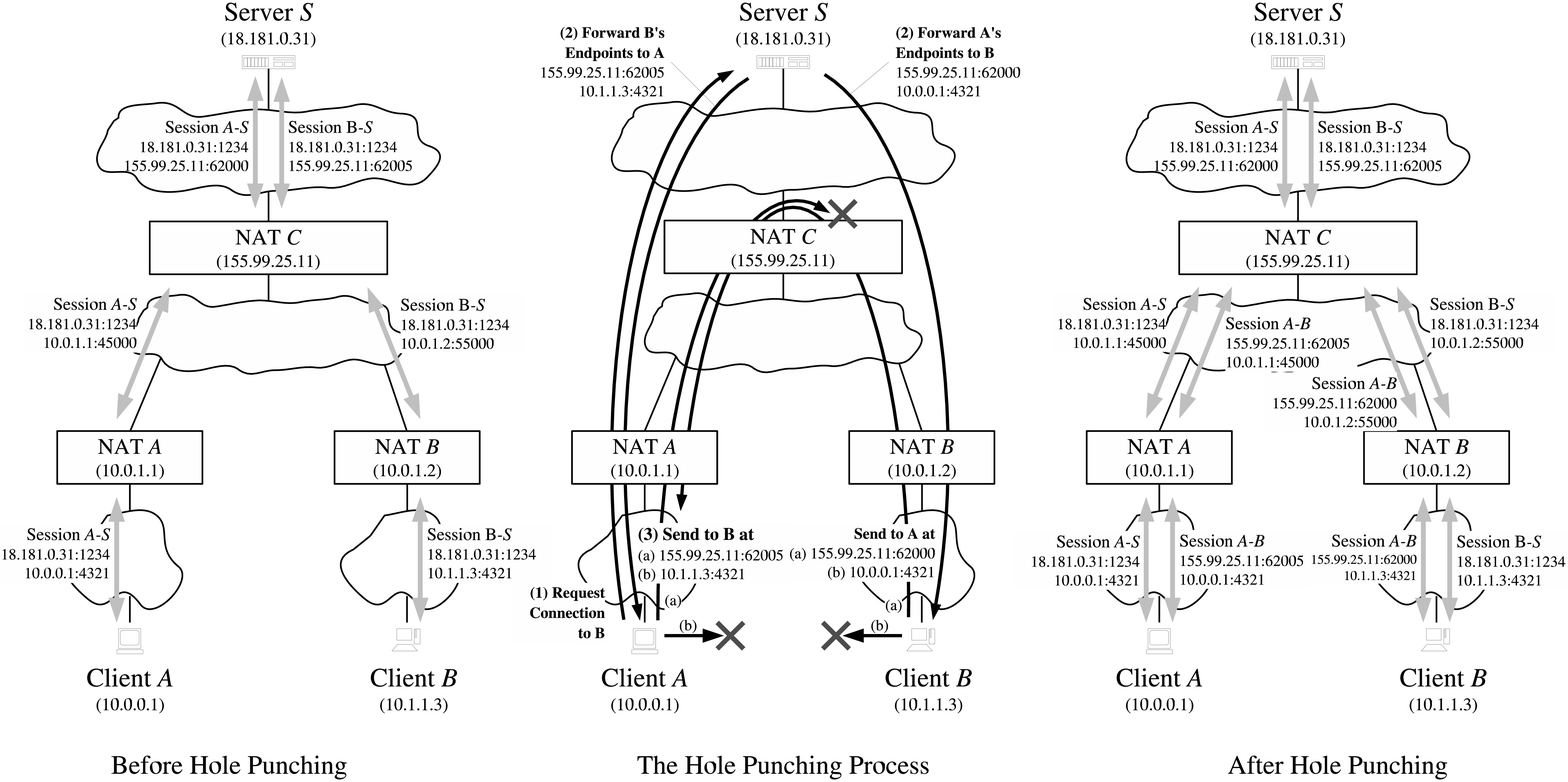, scale=0.34}}
\caption{UDP Hole Punching, Peers Behind Multiple Levels of NAT}
\label{fig-multinat}
\end{figure*}

\subsection{The Rendezvous Server}

Hole punching assumes that the two clients,
$A$ and $B$,
already have active UDP sessions with a rendezvous server $S$.
When a client registers with $S$,
the server records {\em two} endpoints for that client:
the (IP address, UDP port) pair
that the client {\em believes} itself to be using to talk with $S$,
and the (IP address, UDP port) pair
that the server {\em observes} the client to be using to talk with it.
We refer to the first pair as the client's {\em private} endpoint
and the second as the client's {\em public} endpoint.
The server might
obtain the client's private endpoint from the client itself
in a field in the body of the client's registration message,
and obtain the client's public endpoint
from the source IP address and source UDP port fields
in the IP and UDP headers of that registration message.
If the client is {\em not} behind a NAT,
then its private and public endpoints should be identical.

A few poorly behaved NATs are known to scan the body of UDP datagrams
for 4-byte fields that look like IP addresses,
and translate them as they would the IP address fields
in the IP header.
To be robust against such behavior,
applications may wish to obfuscate IP addresses in messages bodies slightly,
for example by transmitting the one's complement of the IP address
instead of the IP address itself.
Of course, if the application is encrypting its messages,
then this behavior is not likely to be a problem.

\subsection{Establishing Peer-to-Peer Sessions}

Suppose client $A$ wants to establish a UDP
session directly with client $B$.
Hole punching proceeds as follows:

\begin{enumerate}
\item	$A$ initially does not know how to reach $B$,
	so $A$ asks $S$
	for help establishing a UDP session with $B$.

\item	$S$ replies to $A$ with a message
	containing $B$'s public {\em and} private endpoints.
	At the same time,
	$S$ uses its UDP session with $B$
	to send $B$ a connection request message
	containing $A$'s public and private endpoints.
	Once these messages are received,
	$A$ and $B$ know each other's public and private endpoints.

\item	When $A$ receives $B$'s public and private endpoints from $S$,
	$A$ starts sending UDP packets to
	{\em both} of these endpoints,
	and subsequently ``locks in'' whichever endpoint
	first elicits a valid response from $B$.
	Similarly, when $B$ receives $A$'s public and private endpoints
	in the forwarded connection request,
	$B$ starts sending UDP packets to $A$
	at each of $A$'s known endpoints,
	locking in the first endpoint that works.
	The order and timing of these messages are not critical
	as long as they are asynchronous.
\end{enumerate}

We now consider
how UDP hole punching handles
each of three specific network scenarios.
In the first situation,
representing the ``easy'' case,
the two clients actually reside behind the same NAT,
on one private network.
In the second,
most common case,
the clients reside behind different NATs.
In the third scenario,
the clients each reside behind {\em two} levels of NAT:
a common ``first-level'' NAT
deployed by an ISP for example,
and distinct ``second-level'' NATs
such as consumer NAT routers for home networks.

It is in general difficult or impossible for the application itself
to determine the exact physical layout of the network,
and thus which of these scenarios (or the many other possible ones)
actually applies at a given time.
Protocols such as STUN~\cite{rfc3489}
can provide some information about the NATs present on a communication path,
but this information may not always be complete or reliable,
especially when multiple levels of NAT are involved.
Nevertheless,
hole punching works automatically in all of these scenarios
{\em without} the application having to know
the specific network organization,
as long as the NATs involved behave in a reasonable fashion.
(``Reasonable'' behavior for NATs will be described
later in Section~\ref{sec-good}.)

\subsection{Peers Behind a Common NAT}

First consider the simple scenario in which the two clients
(probably unknowingly) happen to reside behind the same NAT,
and are therefore located in the same private IP address realm,
as shown in Figure~\ref{fig-samenat}. 
Client $A$ has established a UDP session with server $S$, to which the
common NAT has assigned its own public port number 62000.
Client $B$ has similarly established a session with $S$,
to which the NAT has assigned public port number 62005.

Suppose that client $A$ uses the hole punching technique outlined above
to establish a UDP session with $B$,
using server $S$ as an introducer.
Client $A$ sends $S$ a message requesting a connection to $B$.
$S$ responds to $A$ with $B$'s public and private endpoints,
and also forwards $A$'s public and private endpoints to $B$.
Both clients then attempt to send UDP datagrams to each other
directly at each of these endpoints.
The messages directed to the public endpoints
may or may not reach their destination,
depending on whether or not the NAT supports hairpin translation
as described below in Section~\ref{sec-hairpin}.
The messages directed at the private endpoints
{\em do} reach their destinations, however,
and since this direct route through the private network
is likely to be faster than an indirect route through the NAT anyway,
the clients are most likely to select the private endpoints
for subsequent regular communication.

By assuming that NATs support hairpin translation,
the application might dispense with the complexity
of trying private as well as public endpoints,
at the cost of making local communication behind a common NAT
unnecessarily pass through the NAT.
As our results in Section~\ref{sec-eval} show, however,
hairpin translation is still much less common among existing NATs
than are other ``P2P-friendly'' NAT behaviors.
For now, therefore,
applications may benefit substantially
by using both public and private endpoints.

\subsection{Peers Behind Different NATs}

Suppose clients $A$ and $B$ have private IP addresses
behind different NATs,
as shown in Figure~\ref{fig-diffnat}.
$A$ and $B$ have each initiated UDP communication
sessions from their local port 4321
to port 1234 on server $S$.
In handling these outbound sessions,
NAT $A$ has assigned port 62000 at its own public IP address, 155.99.25.11,
for the use of $A$'s session with $S$,
and NAT $B$ has assigned port 31000 at its IP address, 138.76.29.7,
to $B$'s session with $S$.

In $A$'s registration message to $S$,
$A$ reports its private endpoint to $S$ as 10.0.0.1:4321,
where 10.0.0.1 is $A$'s IP address on its own private network.
$S$ records $A$'s reported private endpoint,
along with $A$'s public endpoint as observed by $S$ itself.
$A$'s public endpoint in this case is 155.99.25.11:62000,
the temporary endpoint assigned to the session by the NAT.
Similarly, when client $B$ registers,
$S$ records $B$'s private endpoint as 10.1.1.3:4321
and $B$'s public endpoint as 138.76.29.7:31000.

Now client $A$ follows the hole punching procedure described above
to establish a UDP communication session directly with $B$.
First, $A$ sends a request message to $S$
asking for help connecting with $B$.
In response, $S$ sends $B$'s public and private endpoints to $A$,
and sends $A$'s public and private endpoints to $B$.
$A$ and $B$ each start trying to send UDP datagrams
directly to each of these endpoints.

Since $A$ and $B$ are on different private networks
and their respective private IP addresses are not globally routable,
the messages sent to these endpoints
will reach either the wrong host or no host at all.
Because many NATs also act as DHCP servers,
handing out IP addresses in a fairly deterministic way
from a private address pool usually determined by the NAT vendor by default,
it is quite likely in practice
that $A$'s messages directed at $B$'s private endpoint
will reach {\em some} (incorrect) host on $A$'s private network
that happens to have the same private IP address as $B$ does.
Applications must therefore authenticate all messages
in some way to filter out such stray traffic robustly.
The messages might include application-specific names or cryptographic tokens,
for example,
or at least a random nonce pre-arranged through $S$.

Now consider $A$'s first message sent to $B$'s public endpoint,
as shown in Figure~\ref{fig-diffnat}.
As this outbound message passes through $A$'s NAT,
this NAT notices that this is the first UDP packet in a new outgoing session.
The new session's source endpoint (10.0.0.1:4321)
is the same as that of the existing session between $A$ and $S$,
but its destination endpoint is different.
If NAT $A$ is well-behaved,
it preserves the identity of $A$'s private endpoint,
consistently translating
{\em all} outbound sessions
from private source endpoint 10.0.0.1:4321
to the corresponding public source endpoint 155.99.25.11:62000.
$A$'s first outgoing message to $B$'s public endpoint
thus, in effect,
``punches a hole'' in $A$'s NAT
for a new UDP session
identified by the endpoints
(10.0.0.1:4321, 138.76.29.7:31000) on $A$'s private network,
and by the endpoints
(155.99.25.11:62000, 138.76.29.7:31000) on the main Internet.

If $A$'s message to $B$'s public endpoint
reaches $B$'s NAT before $B$'s first message to $A$
has crossed $B$'s own NAT,
then $B$'s NAT may interpret $A$'s inbound message
as unsolicited incoming traffic and drop it.
$B$'s first message to $A$'s public address, however,
similarly opens a hole in $B$'s NAT,
for a new UDP session
identified by the endpoints
(10.1.1.3:4321, 155.99.25.11:62000) on $B$'s private network,
and by the endpoints
(138.76.29.7:31000, 155.99.25.11:62000) on the Internet.
Once the first messages from $A$ and $B$ have crossed their respective NATs,
holes are open in each direction
and UDP communication can proceed normally.
Once the clients have verified that the public endpoints work,
they can stop sending messages to the alternative private endpoints.

\com{
XXX move this somewhere else?

The UDP hole punching technique has several useful properties. Once
a direct peer-to-peer UDP connection has been established between two
clients behind NAT devices, either party on that connection can in
turn take over the role of ``introducer'' and help the other party
establish peer-to-peer connections with additional peers, minimizing
the load on the initial introduction server $S$.  The application does
not need to attempt to detect the kind of NAT device it is behind, 
if any [STUN], since the procedure above will establish peer-to-peer
communication channels equally well if either or both clients do not
happen to be behind a NAT device. The UDP hole punching technique
even works automatically with multiple NATs, where one or both
clients are removed from the public Internet via two or more levels
of address translation.
}

\subsection{Peers Behind Multiple Levels of NAT}
\label{sec-hairpin}

In some topologies involving multiple NAT devices,
two clients cannot establish an ``optimal'' P2P route between
them without specific knowledge of the topology.
Consider a final scenario,
depicted in Figure~\ref{fig-multinat}.
Suppose NAT $C$ is a large industrial NAT deployed by an internet
service provider (ISP) to multiplex many customers onto a few public
IP addresses, and NATs $A$ and $B$ are small consumer NAT routers
deployed independently by two of the ISP's customers to multiplex
their private home networks onto their respective ISP-provided IP
addresses.  Only server $S$ and NAT $C$ have globally routable IP
addresses; the ``public'' IP addresses used by NAT $A$ and NAT $B$ are
actually private to the ISP's address realm, while client $A$'s and
$B$'s addresses in turn are private to the addressing realms of NAT $A$
and NAT $B$, respectively.
Each client initiates an outgoing connection to server $S$ as before,
causing NATs $A$ and $B$ each to create a single
public/private translation,
and causing NAT $C$ to establish a public/private translation
for each session.

Now suppose $A$ and $B$ attempt to establish
a direct peer-to-peer UDP connection
via hole punching.
The optimal routing strategy would be for client $A$ to
send messages to client $B$'s ``semi-public'' endpoint at NAT $B$,
10.0.1.2:55000 in the ISP's addressing realm,
and for client $B$ to send messages
to $A$'s ``semi-public'' endpoint at NAT $B$,
namely 10.0.1.1:45000.
Unfortunately, $A$ and $B$ have no way to learn these addresses,
because server $S$ only sees the truly global public endpoints of the clients,
155.99.25.11:62000 and 155.99.25.11:62005 respectively.
Even if $A$ and $B$ had some way to learn these addresses,
there is still no guarantee that they would be usable,
because the address assignments in the ISP's private address realm
might conflict with unrelated address assignments
in the clients' private realms.
(NAT $A$'s IP address in NAT $C$'s realm
might just as easily have been 10.1.1.3, for example,
the same as client $B$'s private address in NAT $B$'s realm.)

The clients
therefore have no choice
but to use their global public addresses as
seen by $S$ for their P2P communication,
and rely on NAT $C$ providing {\em hairpin} or {\em loopback} translation.
When $A$ sends a UDP datagram to $B$'s global endpoint,
155.99.25.11:62005,
NAT $A$ first translates the datagram's source endpoint
from 10.0.0.1:4321 to 10.0.1.1:45000.
The datagram now reaches NAT $C$,
which recognizes that the datagram's destination address
is one of NAT $C$'s own translated {\em public} endpoints.
If NAT $C$ is well-behaved,
it then translates {\em both}
the source and destination addresses in the datagram
and ``loops'' the datagram back onto the private network,
now with a source endpoint of 155.99.25.11:62000
and a destination endpoint of 10.0.1.2:55000.
NAT $B$ finally translates the datagram's destination address
as the datagram enters $B$'s private network,
and the datagram reaches $B$.
The path back to $A$ works similarly.
Many NATs do not yet support hairpin translation,
but it is becoming more common
as NAT vendors become aware of this issue.

\com{
\subsubsection{Assumption of P2P-friendly NAT devices enroute}

The UDP hole punching technique has a caveat in that it works only
if the traversing NAT is a P2P-friendly NAT, such as a Cone NAT.
When a symmetric NAT is encountered enroute, P2P application is 
unable to reuse an already-established translation endpoint for
communication with different external destinations and the
technique would fail. However, Cone NATs are widely deployed in
the Internet. That makes the UDP hole punching technique broadly
applicable; nevertheless a substantial fraction of deployed NATs
are symmetric NATs and do not support the UDP hole punching
technique.
}

\subsection{UDP Idle Timeouts}

Since the UDP transport protocol provides NATs
with no reliable, application-independent way
to determine the lifetime of a session crossing the NAT,
most NATs simply associate an idle timer with UDP translations,
closing the hole if no traffic has used it for some time period.
There is unfortunately no standard value for this timer:
some NATs have timeouts as short as 20 seconds.
If the application needs to keep an idle UDP session active
after establishing the session via hole punching,
the application must send periodic keep-alive packets
to ensure that the relevant translation state in the NATs
does not disappear.

Unfortunately,
many NATs associate UDP idle timers with individual UDP sessions
defined by a particular pair of endpoints,
so sending keep-alives on one session
will not keep other sessions active
even if all the sessions originate from
the same private endpoint.
Instead of sending keep-alives on many different P2P sessions,
applications can avoid excessive keep-alive traffic
by detecting when a UDP session no longer works,
and re-running the original hole punching procedure again ``on demand.''

%% file: tcp.tex
\section{TCP Hole Punching}
\label{sec-tcp}

Establishing peer-to-peer TCP connections
between hosts behind NATs
is slightly more complex than for UDP,
but TCP hole punching is remarkably similar
at the protocol level.
Since it is not as well-understood,
it is currently supported by fewer existing NATs.
When the NATs involved {\em do} support it, however,
TCP hole punching is just as fast and reliable as UDP hole punching.
Peer-to-peer TCP communication across well-behaved NATs
may in fact be {\em more} robust than UDP communication,
because unlike UDP,
the TCP protocol's state machine
gives NATs on the path a standard way to determine
the precise lifetime of a particular TCP session.

\subsection{Sockets and TCP Port Reuse}

The main practical challenge
to applications wishing to implement TCP hole punching
is not a protocol issue
but an application programming interface (API) issue.
Because the standard Berkeley sockets API
was designed around the client/server paradigm,
the API allows
a TCP stream socket
to be used to initiate an outgoing connection via {\tt connect()},
or to listen for incoming connections via {\tt listen()} and {\tt accept()},
{\em but not both}.
Further,
TCP sockets usually have a one-to-one correspondence
to TCP port numbers on the local host:
after the application binds one socket to a particular local TCP port,
attempts to bind a second socket to the same TCP port fail.

For TCP hole punching to work, however,
we need to use a single local TCP port
to listen for incoming TCP connections
and to initiate multiple outgoing TCP connections concurrently.
Fortunately, all major operating systems
support a special TCP socket option,
commonly named \verb|SO_REUSEADDR|,
which allows the application to bind multiple sockets
to the same local endpoint
as long as this option is set on all of the sockets involved.
BSD systems have introduced a \verb|SO_REUSEPORT| option
that controls port reuse separately from address reuse;
on such systems {\em both} of these options must be set.

\subsection{Opening Peer-to-Peer TCP Streams}
\label{sec-tcp-steps}

Suppose that client $A$ wishes to set up a TCP connection with client $B$.
We assume as usual
that both $A$ and $B$ already have active TCP connections
with a well-known rendezvous server $S$.
The server records
each registered client's public and private endpoints,
just as for UDP.
At the protocol level,
TCP hole punching works almost exactly as for UDP:

\begin{enumerate}
\item	Client $A$ uses its active TCP session with $S$
	to ask $S$ for help connecting to $B$.

\item	$S$ replies to $A$ with $B$'s public and private TCP endpoints,
	and at the same time sends $A$'s public and private endpoints to $B$.

\item	From {\em the same local TCP ports}
	that $A$ and $B$ used to register with $S$,
	$A$ and $B$ each asynchronously make outgoing connection attempts
	to the other's public and private endpoints
	as reported by $S$,
	while simultaneously listening for incoming connections
	on their respective local TCP ports.

\item	$A$ and $B$ wait for outgoing connection attempts to succeed,
	and/or for incoming connections to appear.
	If one of the outgoing connection attempts fails
	due to a network error
	such as ``connection reset'' or ``host unreachable,''
	the host simply re-tries that connection attempt
	after a short delay (e.g., one second),
	up to an application-defind maximum timeout period.

\item	When a TCP connection is made,
	the hosts authenticate each other
	to verify that they connected to the intended host.
	If authentication fails,
	the clients close that connection
	and continue waiting for others to succeed.
	The clients
	use the first successfully authenticated TCP stream
	resulting from this process.
\end{enumerate}

\begin{figure}[t]
\centerline{\epsfig{file=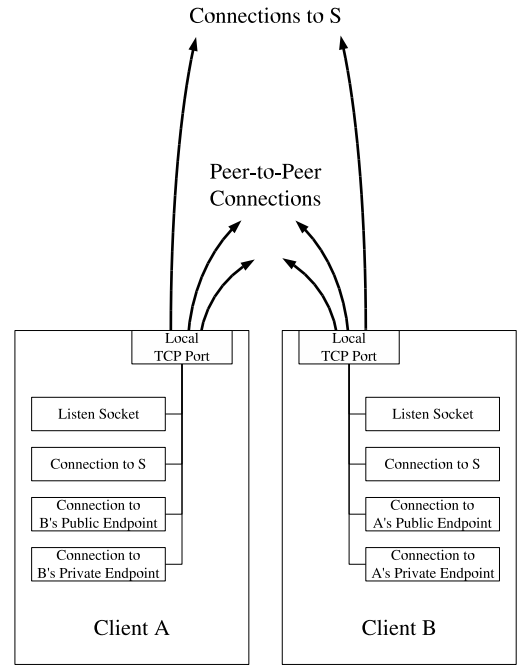, scale=0.35}}
\caption{Sockets versus Ports for TCP Hole Punching}
\label{fig-tcpsocks}
\end{figure}

Unlike with UDP,
where each client only needs one socket
to communicate with both $S$ and any number of peers simultaneously,
with TCP each client application must manage several sockets
bound to a single local TCP port on that client node,
as shown in Figure~\ref{fig-tcpsocks}.
Each client needs a stream socket representing its connection to $S$,
a listen socket on which to accept incoming connections from peers,
and at least two additional stream sockets
with which to initiate outgoing connections
to the other peer's public and private TCP endpoints.

Consider the common-case scenario
in which the clients $A$ and $B$ are behind different NATs,
as shown in Figure~\ref{fig-diffnat},
and assume that the port numbers shown in the figure
are now for TCP rather than UDP ports.
The outgoing connection attempts $A$ and $B$ make
to each other's private endpoints
either fail or connect to the wrong host.
As with UDP,
it is important that TCP applications authenticate their peer-to-peer sessions,
due of the likelihood of mistakenly connecting to
a random host on the local network
that happens to have the same private IP address
as the desired host on a remote private network.

The clients' outgoing connection attempts
to each other's {\em public} endpoints,
however, cause the respective NATs to open up new ``holes''
enabling direct TCP communication between $A$ and $B$.
If the NATs are well-behaved,
then a new peer-to-peer TCP stream
automatically forms between them.
If $A$'s first SYN packet to $B$ reaches $B$'s NAT
before $B$'s first SYN packet to $A$ reaches $B$'s NAT, for example,
then $B$'s NAT may interpret $A$'s SYN
as an unsolicited incoming connection attempt
and drop it.
$B$'s first SYN packet to $A$ should subsequently get through,
however,
because $A$'s NAT sees this SYN
as being part of the outbound session to $B$
that $A$'s first SYN had already initiated.

\subsection{Behavior Observed by the Application}

What the client applications observe to happen with their sockets
during TCP hole punching
depends on the timing and the TCP implementations involved.
Suppose that $A$'s first outbound SYN packet to $B$'s public endpoint
is dropped by NAT $B$,
but $B$'s first subsequent SYN packet to $A$'s public endpoint
gets through to $A$ before $A$'s TCP retransmits its SYN.
Depending on the operating system involved,
one of two things may happen:

\begin{itemize}
\item	$A$'s TCP implementation notices that
	the session endpoints for the incoming SYN
	match those of an outbound session $A$ was attempting to initiate.
	$A$'s TCP stack therefore associates this new session
	with the socket that the local application on $A$
	was using to {\tt connect()} to $B$'s public endpoint.
	The application's asynchronous {\tt connect()} call succeeds,
	and nothing happens with the application's listen socket.

	Since the received SYN packet did not include an ACK
	for $A$'s previous outbound SYN,
	$A$'s TCP replies to $B$'s public endpoint
	with a SYN-ACK packet,
	the SYN part being merely a replay
	of $A$'s original outbound SYN,
	using the same sequence number.
	Once $B$'s TCP receives $A$'s SYN-ACK,
	it responds with its own ACK for $A$'s SYN,
	and the TCP session enters the connected state on both ends.

\item	Alternatively,
	$A$'s TCP implementation might instead notice
	that $A$ has an active listen socket on that port
	waiting for incoming connection attempts.
	Since $B$'s SYN looks like an incoming connection attempt,
	$A$'s TCP creates a {\em new} stream socket
	with which to associate the new TCP session,
	and hands this new socket to the application
	via the application's next {\tt accept()} call
	on its listen socket.
	$A$'s TCP then responds to $B$ with a SYN-ACK as above,
	and TCP connection setup proceeds as usual
	for client/server-style connections.

	Since $A$'s prior outbound {\tt connect()} attempt to $B$
	used a combination of source and destination endpoints
	that is now in use by another socket,
	namely the one just returned to the application via {\tt accept()},
	$A$'s asynchronous {\tt connect()} attempt
	must fail at some point,
	typically with an ``address in use'' error.
	The application nevertheless has the working peer-to-peer
	stream socket it needs to communicate with $B$,
	so it ignores this failure.
\end{itemize}

The first behavior above appears to be usual for BSD-based operating systems,
whereas the second behavior appears more common under Linux and Windows.

\subsection{Simultaneous TCP Open}
\label{sec-tcp-simul}

Suppose that the timing of the various connection attempts
during the hole punching process
works out so that
the initial outgoing SYN packets from {\em both} clients
traverse their respective local NATs,
opening new outbound TCP sessions in each NAT,
before reaching the remote NAT.
In this ``lucky'' case,
the NATs do not reject either of the initial SYN packets,
and the SYNs cross on the wire between the two NATs.
In this case,
the clients observe an event
known as a {\em simultaneous TCP open}:
each peer's TCP receives a ``raw'' SYN
while waiting for a SYN-ACK.
Each peer's TCP responds with a SYN-ACK,
whose SYN part essentially ``replays'' the peer's previous outgoing SYN,
and whose ACK part acknowledges the SYN received from the other peer.

What the respective applications observe in this case
again depends on the behavior of the TCP implementations involved,
as described in the previous section.
If {\em both} clients implement the second behavior above,
it may be that
{\em all} of the asynchronous {\tt connect()} calls
made by the application ultimately fail,
but the application running on each client
nevertheless receives a new, working peer-to-peer TCP stream socket
via {\tt accept()}---%
as if this TCP stream had magically ``created itself'' on the wire
and was merely passively accepted at the endpoints!
As long as the application does not care
whether it ultimately receives its peer-to-peer TCP sockets
via {\tt connect()} or {\tt accept()},
the process results in a working stream
on any TCP implementation
that properly implements the standard TCP state machine
specified in RFC 793~\cite{rfc793}.


Each of the alternative network organization scenarios
discussed in Section~\ref{sec-udp} for UDP
works in exactly the same way for TCP.
For example,
TCP hole punching works in multi-level NAT scenarios
such as the one in Figure~\ref{fig-multinat}
as long as the NATs involved are well-behaved.

\subsection{Sequential Hole Punching}
\label{sec-tcp-seq}

In a variant of the above TCP hole punching procedure
implemented by the NatTrav library~\cite{eppinger05tcp},
the clients attempt connections to each other
sequentially rather than in parallel.
For example:
(1) $A$ informs $B$ via $S$ of its desire to communicate,
{\em without} simultaneously listening on its local port;
(2) $B$ makes a {\tt connect()} attempt to $A$,
which opens a hole in $B$'s NAT
but then fails due to a timeout
or RST from $A$'s NAT or a RST from $A$ itself;
(3) $B$ closes its connection to $S$
and does a {\tt listen()} on its local port;
(4) $S$ in turn closes its connection with $A$,
signaling $A$ to attempt a {\tt connect()} directly to $B$.

This sequential procedure may be particularly useful
on Windows hosts prior to XP Service Pack 2,
which did not correctly implement simultaneous TCP open,
or on sockets APIs
that do not support the \verb|SO_REUSEADDR| functionality.
The sequential procedure is more timing-dependent,
however,
and may be slower in the common case
and less robust in unusual situations.
In step (2), for example,
$B$ must allow its ``doomed-to-fail'' {\tt connect()} attempt
enough time to ensure that at least one SYN packet
traverses all NATs on its side of the network.
Too little delay risks
a lost SYN derailing the process,
whereas too much delay increases the total time
required for hole punching.
The sequential hole punching procedure also effectively ``consumes''
both clients' connections to the server $S$,
requiring the clients to open fresh connections to $S$
for each new P2P connection to be forged.
The parallel hole punching procedure,
in contrast,
typically completes as soon as
both clients make their outgoing {\tt connect()} attempts,
and allows each client to retain and re-use
a single connection to $S$ indefinitely.

%% file: good.tex
\section{Properties of P2P-Friendly NATs}
\label{sec-good}

This section describes the key behavioral properties NATs must have
in order for the hole punching techniques described above to work properly.
Not all current NAT implementations satisfy these properties,
but many do,
and NATs are gradually becoming more ``P2P-friendly''
as NAT vendors recognize the demand for peer-to-peer protocols
such as voice over IP and on-line gaming.

This section is not meant to be
a complete or definitive specification for how NATs ``should'' behave;
we provide it merely for information
about the most commonly observed behaviors
that enable or break P2P hole punching.
The IETF has started a new working group, BEHAVE,
to define official ``best current practices'' for NAT behavior.
The BEHAVE group's initial drafts
include the considerations outlined in this section and others;
NAT vendors should of course
follow the IETF working group directly
as official behavioral standards are formulated.


\subsection{Consistent Endpoint Translation}
\label{sec-good-consist}

The hole punching techniques described here
only work automatically
if the NAT consistently maps a given TCP or UDP source endpoint
on the private network
to a {\em single} corresponding public endpoint
controlled by the NAT.
A NAT that behaves in this way
is referred to as a {\em cone NAT}
in RFC 3489~\cite{rfc3489} and elsewhere,
because the NAT ``focuses'' all sessions 
originating from a single private endpoint
through the same public endpoint on the NAT.

Consider again the scenario in Figure~\ref{fig-diffnat}, for example.
When client $A$ initially contacted the well-known server $S$,
NAT $A$ chose to use port 62000
at its own public IP address, 155.99.25.11,
as a temporary public endpoint
to representing $A$'s private endpoint 10.0.0.1:4321.
When $A$ later attempts to establish a peer-to-peer session with $B$
by sending a message from the same local private endpoint
to $B$'s public endpoint,
$A$ depends on NAT $A$ preserving the identity of this private endpoint,
and re-using the existing public endpoint of 155.99.25.11:62000,
because that is the public endpoint for $A$
to which $B$ will be sending its corresponding messages.

A NAT that is only designed to support client/server protocols
will not necessarily preserve the identities of private endpoints in this way.
Such a NAT is a {\em symmetric NAT} in RFC 3489 terminology.
For example,
after the NAT assigns the public endpoint 155.99.25.11:62000
to client $A$'s session with server $S$,
the NAT might assign a different public endpoint,
such as 155.99.25.11:62001,
to the P2P session that $A$ tries to initiate with $B$.
In this case,
the hole punching process fails to provide connectivity,
because the subsequent incoming messages from $B$
reach NAT $A$ at the wrong port number.

Many symmetric NATs allocate port numbers for successive sessions
in a fairly predictable way.
Exploiting this fact,
variants of hole punching algorithms~\cite{guha04nutss, biggadike05natblaster}
can be made to work ``much of the time'' even over symmetric NATs
by first probing the NAT's behavior
using a protocol such as STUN~\cite{rfc3489},
and using the resulting information
to ``predict'' the public port number the NAT
will assign to a new session.
Such prediction techniques amount to chasing a moving target, however,
and many things can go wrong along the way.
The predicted port number might already be in use
causing the NAT to jump to another port number,
for example,
or another client behind the same NAT
might initiate an unrelated session at the wrong time
so as to allocate the predicted port number.
While port number prediction can be a useful trick
for achieving maximum compatibility with badly-behaved existing NATs,
it does not represent a robust long-term solution.
Since symmetric NAT provides no greater security
than a cone NAT with per-session traffic filtering,
symmetric NAT is becoming less common
as NAT vendors adapt their algorithms to support P2P protocols.

\subsection{Handling Unsolicited TCP Connections}
\label{sec-good-tcp}

When a NAT receives a SYN packet on its public side
for what appears to be an unsolicited incoming connection attempt,
it is important that the NAT just silently drop the SYN packet.
Some NATs instead actively reject such incoming connections
by sending back a TCP RST packet or even an ICMP error report,
which interferes with the TCP hole punching process.
Such behavior is not necessarily fatal,
as long as the applications re-try outgoing connection attempts
as specified in step 4 of the process described in Section~\ref{sec-tcp-steps},
but the resulting transient errors
can make hole punching take longer.

\com{
Consider the scenario in Figure~\ref{fig-diffnat},
for example.
Suppose client $A$'s first SYN packet directed at $B$'s public endpoint
reaches NAT $B$ before client $B$'s first SYN packet has reached NAT $B$
and opened up a TCP hole.
NAT $B$ therefore interprets this SYN
as an unsolicited incoming connection attempt.
If NAT $B$ sends an RST back to $A$ instead of simply dropping the SYN,
then this RST may in turn cause NAT $A$
to close the hole that $A$'s outbound SYN just opened.
When $B$'s first SYN packet to $A$ subsequently opens a hole in NAT $B$
and arrives at NAT $A$,
$B$'s SYN is then rejected as an unsolicited incoming connection by NAT $A$.
If NAT $A$ also responds with a RST,
NAT $B$ may in turn close its temporary hole as well,
and no progress has been made.

Even when both NATs reject unsolicited incoming connection attempts
with RST packets,
it is still possible to establish P2P TCP streams
if $A$'s and $B$'s outgoing SYN packets
can be timed so that each client's SYN traverses its local NAT
before reaching the remote NAT.
In this case,
the ``lucky'' scenario described in Section~\ref{sec-tcp-simul}
is the only one that will work at all,
and hole punching becomes timing-dependent and therefore much less robust.
}

\subsection{Leaving Payloads Alone}

A few existing NATs
are known to scan ``blindly'' through packet payloads
for 4-byte values that look like IP addresses,
and translate them as they would the IP address in the packet header,
without knowing anything about the application protocol in use.
This bad behavior fortunately appears to be uncommon,
and applications can easily protect themselves against it
by obfuscating IP addresses they send in messages,
for example by sending the bitwise complement of the desired IP address.

\subsection{Hairpin Translation}
\label{sec-good-hairpin}

Some multi-level NAT situations
require hairpin translation support
in order for either TCP or UDP hole punching to work,
as described in Section~\ref{sec-hairpin}.
The scenario shown in Figure~\ref{fig-multinat}, for example,
depends on NAT $C$ providing hairpin translation.
Support for hairpin translation
is unfortunately rare in current NATs,
but fortunately so are the network scenarios that require it.
Multi-level NAT is becoming more common
as IPv4 address space depletion continues,
however,
so support for hairpin translation
is important in future NAT implementations.


%% file: eval.tex
\section{Evaluation of Existing NATs}
\label{sec-eval}

To evaluate the robustness of the TCP and UDP hole punching techniques
described in this paper
on a variety of existing NATs,
we implemented and distributed a test program called
NAT Check~\cite{natcheck},
and solicited data from Internet users about their NATs.

NAT Check's primary purpose
is to test NATs
for the two behavioral properties most crucial
to reliable UDP and TCP hole punching:
namely,
consistent identity-preserving endpoint translation
(Section~\ref{sec-good-consist}),
and silently dropping unsolicited incoming TCP SYNs
instead of rejecting them with RSTs or ICMP errors
(Section~\ref{sec-good-tcp}).
In addition,
NAT Check separately tests
whether the NAT supports hairpin translation
(Section~\ref{sec-good-hairpin}),
and whether the NAT filters unsolicited incoming traffic at all.
This last property does not affect hole punching,
but provides a useful indication the NAT's firewall policy.

NAT Check makes no attempt
to test every relevant facet of NAT behavior individually:
a wide variety of subtle behavioral differences are known,
some of which are difficult
to test reliably~\cite{jennings04nat}.
Instead, NAT Check merely attempts to answer the question,
``how commonly can the proposed hole punching techniques
be expected to work on deployed NATs,
under typical network conditions?''

\subsection{Test Method}

NAT Check consists of a client program
to be run on a machine behind the NAT to be tested,
and three well-known servers at different global IP addresses.
The client cooperates with the three servers
to check the NAT behavior
relevant to both TCP and UDP hole punching.
The client program is small and relatively portable,
currently running on Windows, Linux, BSD, and Mac OS X.
The machines hosting the well-known servers all run FreeBSD.

\subsubsection{UDP Test}

\begin{figure}[t]
\centerline{\epsfig{file=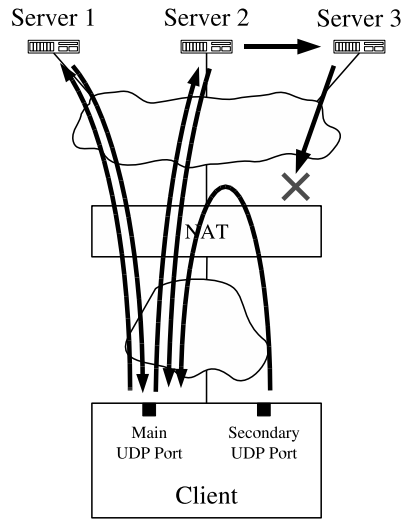, scale=0.40}}
\caption{NAT Check Test Method for UDP}
\label{fig-udptest}
\end{figure}

To test the NAT's behavior for UDP,
the client opens a socket and binds it to a local UDP port,
then successively sends ``ping''-like requests to servers 1 and 2,
as shown in Figure~\ref{fig-udptest}.
These servers each respond to the client's pings
with a reply that includes the client's public UDP endpoint:
the client's own IP address and UDP port number
as observed by the server.
If the two servers report the same public endpoint for the client,
NAT Check assumes that the NAT properly preserves
the identity of the client's private endpoint,
satisfying the primary precondition for reliable UDP hole punching.

When server 2 receives a UDP request from the client,
besides replying directly to the client
it also forwards the request to server 3,
which in turn replies to the client
from its own IP address.
If the NAT's firewall properly filters
``unsolicited'' incoming traffic on a per-session basis,
then the client never sees these replies from server 3,
even though they are directed at the same public port
as the replies from servers 1 and 2.

To test the NAT for hairpin translation support,
the client simply opens a second UDP socket at a different local port
and uses it to send messages to the {\em public} endpoint
representing the client's first UDP socket,
as reported by server 2.
If these messages reach the client's first private endpoint,
then the NAT supports hairpin translation.

\subsubsection{TCP Test}

The TCP test follows a similar pattern as for UDP.
The client uses a single local TCP port
to initiate outbound sessions to servers 1 and 2,
and checks whether the public endpoints
reported by servers 1 and 2 are the same,
the first precondition
for reliable TCP hole punching.

The NAT's response to unsolicited incoming connection attempts
also impacts the speed and reliability of TCP hole punching, however,
so NAT Check also tests this behavior.
When server 2 receives the client's request,
instead of immediately replying to the client,
it forwards a request to server 3
and waits for server 3 to respond with a ``go-ahead'' signal.
When server 3 receives this forwarded request,
it attempts to initiate an inbound connection
to the client's public TCP endpoint.
Server 3 waits up to five seconds 
for this connection to succeed or fail,
and if the connection attempt is still ``in progress'' after five seconds,
server 3 responds to server 2 with the ``go-ahead'' signal
and continues waiting for up to 20 seconds.
Once the client finally receives server 2's reply
(which server 2 delayed waiting for server 3's ``go-ahead'' signal),
the client attempts an outbound connection to server 3,
effectively causing a simultaneous TCP open with server 3.

What happens during this test
depends on the NAT's behavior as follows.
If the NAT properly just drops server 3's ``unsolicited'' incoming SYN packets,
then nothing happens on the client's listen socket
during the five second period
before server 2 replies to the client.
When the client finally initiates its own connection to server 3,
opening a hole through the NAT,
the attempt succeeds immediately.
If on the other hand
the NAT does {\em not} drop server 3's unsolicited incoming SYNs
but allows them through
(which is fine for hole punching but not ideal for security),
then the client receives an incoming TCP connection on its listen socket
before receiving server 2's reply.
Finally, if the NAT actively rejects server 3's unsolicited incoming SYNs
by sending back TCP RST packets,
then server 3 gives up
and the client's subsequent attempt to connect to server 3 fails.

To test hairpin translation for TCP,
the client simply uses a secondary local TCP port
to attempt a connection to the public endpoint
corresponding to its primary TCP port,
in the same way as for UDP.

\subsection{Test Results}


\begin{table*}
\begin{center}
\begin{tabular}{ll|rr|rr|rr|rr|}
&		& \multicolumn{4}{|c|}{\bf UDP}
					& \multicolumn{4}{|c|}{\bf TCP} \\
&		& \multicolumn{2}{|c}{Hole}
				& \multicolumn{2}{c|}{}
			& \multicolumn{2}{|c}{Hole}
					& \multicolumn{2}{c|}{} \\
&		& \multicolumn{2}{|c|}{Punching}
				& \multicolumn{2}{|c|}{Hairpin}
			& \multicolumn{2}{|c|}{Punching}
					& \multicolumn{2}{|c|}{Hairpin} \\
\hline
\multicolumn{2}{l|}{\bf NAT Hardware} &	& & & & & & & \\
& Linksys	&45/46&(98\%)	&5/42&(12\%)	&33/38&(87\%)	&3/38&(8\%)\\
& Netgear	&31/37&(84\%)	&3/35&(9\%)	&19/30&(63\%)	&0/30&(0\%)\\
& D-Link	&16/21&(76\%)	&11/21&(52\%)	&9/19&(47\%)	&2/19&(11\%)\\
& Draytek	&2/17&(12\%)	&3/12&(25\%)	&2/7&(29\%)	&0/7&(0\%)\\
& Belkin	&14/14&(100\%)	&1/14&(7\%)	&11/11&(100\%)	&0/11&(0\%)\\
& Cisco		&12/12&(100\%)	&3/9&(33\%)	&6/7&(86\%)	&2/7&(29\%)\\
& SMC		&12/12&(100\%)	&3/10&(30\%)	&8/9&(89\%)	&2/9&(22\%)\\
& ZyXEL		&7/9&(78\%)	&1/8&(13\%)	&0/7&(0\%)	&0/7&(0\%)\\
& 3Com		&7/7&(100\%)	&1/7&(14\%)	&5/6&(83\%)	&0/6&(0\%)\\
\hline
\multicolumn{2}{l|}{\bf OS-based NAT} &	& & & & & & & \\
& Windows	&31/33&(94\%)	&11/32&(34\%)	&16/31&(52\%)	&28/31&(90\%)\\
& Linux		&26/32&(81\%)	&3/25&(12\%)	&16/24&(67\%)	&2/24&(8\%)\\
& FreeBSD	&7/9&(78\%)	&3/6&(50\%)	&2/3&(67\%)	&1/1&(100\%)\\
\hline
\multicolumn{2}{l|}{\bf All Vendors}
		&310/380&(82\%)	&80/335&(24\%)	&184/286&(64\%)	&37/286&(13\%) \\
\hline
\end{tabular}
\end{center}
\caption{User Reports of NAT Support for UDP and TCP Hole Punching}
\label{tab-nat}
\end{table*}

The NAT Check data we gathered
consists of 380 reported data points
covering a variety of NAT router hardware from 68 vendors,
as well as the NAT functionality built
into different versions of eight popular operating systems.
Only 335 of the total data points include results for UDP hairpin translation,
and only 286 data points include results for TCP,
because we implemented these features in later versions of NAT Check
after we had already started gathering results.
The data is summarized by NAT vendor in Table~\ref{tab-nat};
the table only individually lists vendors
for which at least five data points were available.
The variations in the test results for a given vendor
can be accounted for by a variety of factors,
such as different NAT devices or product lines sold by the same vendor,
different software or firmware versions of the same NAT implementation,
different configurations,
and probably occasional NAT Check testing or reporting errors.

Out of the 380 reported data points for UDP,
in 310 cases (82\%)
the NAT consistently translated the client's private endpoint,
indicating basic compatibility with UDP hole punching.
Support for hairpin translation is much less common, however:
of the 335 data points that include UDP hairpin translation results,
only 80 (24\%) show hairpin translation support.

Out of the 286 data points for TCP,
184 (64\%) show compatibility with TCP hole punching:
the NAT consistently translates the client's private TCP endpoint,
and does not send back RST packets
in response to unsolicited incoming connection attempts.
Hairpin translation support is again much less common:
only 37 (13\%) of the reports
showed hairpin support for TCP.

Since these reports 
were generated by a ``self-selecting'' community of volunteers,
they do not constitute a random sample
and thus do not necessarily represent
the true distribution of the NATs in common use.
The results are nevertheless encouraging:
it appears that the majority of commonly-deployed NATs
already support UDP and TCP hole punching
at least in single-level NAT scenarios.

\com{
Vendors:
	@it
	2Wire
	3Com
	A-link
	Actiontec
	Aethra
	Airlink Plus
	Alcatel
	Allied Data
	Apple
	Aramiska
	Arescom
	Astaro
	Asus
	Atlantis
	Aztech
	Belkin
	BenQ
	Billion
	BT Voyager
	Cayman DSL
	Check Point
	Cisco
	Corega
	D-Link
	Dick Smith Electronics
	Draytek
	DSE NZ
	DYNAMODE
	EDIMAX
	Ericsson
	Gnet - GVC
	Ingate
	Intel
	Intertex
	LANCOM
	Linksys
	M0n0wall
	Mentor
	Microsoft
	Motorola
	MultiTech
	NetComm
	Netgear
	Netopia
	NetScreen
	Nokia
	Paradigm
	Parks / Hyundai
	Peak
	Planet
	Planex
	Sagem
	Scientific Atlanta
	Siemens
	Sitecom
	SMC
	SonicWall
	Sweex
	Tecnew
	Telekom Germany
	Telewell
	Thomson/Alcatel
	TopCam
	TP-LINK
	US Robotics
	XSense
	Zyxel

OS's:
	FreeBSD
	Linux
	MS-DOS
	NetBSD
	NetWare
	Mac X
	OpenBSD
	Windows
}

\subsection{Testing Limitations}

There are a few limitations in NAT Check's current testing protocol
that may cause misleading results in some cases.
First, we only learned recently
that a few NAT implementations blindly
translate IP addresses they find in unknown application payloads,
and the NAT Check protocol currently does not protect itself from this behavior
by obfuscating the IP addresses it transmits.

Second,
NAT Check's current hairpin translation checking
may yield unnecessarily pessimistic results
because it does not use the full, two-way hole punching procedure
for this test.
NAT Check currently assumes that a NAT supporting hairpin translation
does not filter ``incoming'' hairpin connections
arriving from the private network
in the way it would filter incoming connections
arriving at the public side of the NAT,
because such filtering is unnecessary for security.
We later realized, however,
that a NAT might simplistically
treat {\em any} traffic directed at the NAT's public ports
as ``untrusted'' regardless of its origin.
We do not yet know which behavior is more common.

Finally,
NAT implementations exist
that consistently translate the client's private endpoint
as long as {\em only one} client behind the NAT
is using a particular private port number,
but switch to symmetric NAT or even worse behaviors
if two or more clients with different IP addresses on the private network
try to communicate through the NAT from the same private port number.
NAT Check could only detect this behavior
by requiring the user to run it on two or more client hosts
behind the NAT at the same time.
Doing so would make NAT Check much more difficult to use, however,
and impossible for users who only have one usable machine behind the NAT.
Nevertheless, we plan to implement this testing functionality
as an option in a future version of NAT Check.

\subsection{Corroboration of Results}

Despite testing difficulties such as those above,
our results are generally corroborated
by those of a large ISP,
who recently found that of the top three consumer NAT router vendors,
representing 86\% of the NATs observed on their network,
all three vendors currently produce NATs compatible
with UDP hole punching~\cite{uberti04}.
Additional independent results
recently obtained
using the UDP-oriented STUN protocol~\cite{jennings04nat},
and STUNT, a TCP-enabled extension~\cite{guha-stunt, guha04nutss},
also appear consistent with
our results.
These latter studies provide more information on each NAT
by testing a wider variety of behaviors individually,
instead of just testing for basic hole punching compatibility
as NAT Check does.
Since these more extensive tests
require multiple cooperating clients behind the NAT
and thus are more difficult to run,
however,
these results are so far available
on a more limited variety of NATs.

%% file: related.tex
\section{Related Work}
\label{sec-related}

UDP hole punching
was first explored and publicly documented by Dan Kegel~\cite{kegel99nat},
and is by now well-known in peer-to-peer application communities.
Important aspects of UDP hole punching have also been indirectly documented
in the specifications of several experimental protocols,
such as STUN~\cite{rfc3489},
ICE~\cite{rosenberg03ice},
and Teredo~\cite{huitema04teredo}.
We know of no existing published work
that thoroughly analyzes hole punching,
however,
or that points out the hairpin translation issue
for multi-level NAT (Section~\ref{sec-hairpin}).

We also know of no prior work
that develops TCP hole punching
in the symmetric fashion described here.
Even the existence of the crucial
\verb|SO_REUSEADDR|/\linebreak[0]\verb|SO_REUSEPORT| options
in the Berkeley sockets API
appears to be little-known among P2P application developers.
NatTrav~\cite{eppinger05tcp}
implements a similar but asymmetric TCP hole punching procedure
outlined earlier in Section~\ref{sec-tcp-seq}.
NUTSS~\cite{guha04nutss}
and NATBLASTER~\cite{biggadike05natblaster}
implement more complex TCP hole punching tricks
that can work around some of the bad NAT behaviors
mentioned in Section~\ref{sec-good},
but they require the rendezvous server to spoof source IP addresses,
and they also require the client applications to have access to ``raw'' sockets,
usually available only at root or administrator privilege levels.

Protocols such as SOCKS~\cite{rfc1928},
UPnP~\cite{upnp01igd},
and MIDCOM~\cite{rfc3303}
allow applications to traverse a NAT
through explicit cooperation with the NAT.
These protocols are not widely or consistently supported
by NAT vendors or applications, however,
and do not appear to address
the increasingly important multi-level NAT scenarios.
Explicit control of a NAT further
requires the application to locate the NAT
and perhaps authenticate itself,
which typically involves explicit user configuration.
When hole punching works,
in contrast,
it works with no user intervention.

Recent proposals
such as HIP~\cite{moskowitz03hip-arch}
and FARA~\cite{clark03fara}
extend the Internet's basic architecture
by decoupling a host's identity from its location~\cite{saltzer82naming}.
IPNL~\cite{francis02ipnl},
UIP~\cite{ford03scalable, ford03uip},
and DOA~\cite{walfish04middleboxes}
propose schemes for routing across NATs
in such an architecture.
While such extensions
are probably needed in the long term,
hole punching enables applications
to work over the existing network infrastructure immediately
with no protocol stack upgrades,
and leaves the notion of ``host identity'' for applications to define.

%% file: conc.tex
\section{Conclusion}
\label{sec-conc}

Hole punching is a general-purpose technique
for establishing peer-to-peer connections in the presence of NAT.
As long as the NATs involved meet certain behavioral requirements,
hole punching works consistently and robustly
for both TCP and UDP communication,
and can be implemented by ordinary applications
with no special privileges
or specific network topology information.
Hole punching fully preserves the transparency
that is one of the most important hallmarks and attractions of NAT,
and works even with multiple levels of NAT---%
though certain corner case situations require hairpin translation,
a NAT feature not yet widely implemented.

\subsection*{Acknowledgments}

The authors wish to thank Dave Andersen for his crucial support
in gathering the results presented in Section~\ref{sec-eval}.
We also wish to thank Henrik Nordstrom, Christian Huitema,
Justin Uberti, Mema Roussopoulos,
and the anonymous USENIX reviewers
for valuable feedback
on early drafts of this paper.
Finally, we wish to thank the many volunteers
who took the time to run NAT Check on their systems
and submit the results.